\ifdefined\drafting
\documentclass[aps,prb,superscriptaddress, floatfix,citeautoscript,notitlepage,a4paper,longbibliography]{revtex4-2}
\else
\documentclass[aps,prb,twocolumn, superscriptaddress,floatfix, citeautoscript,a4paper,longbibliography]{revtex4-2}
\fi

\usepackage[utf8]{inputenc}
\usepackage[english]{babel}
\usepackage{ragged2e}
\usepackage[final]{graphicx}
\usepackage[centertags, sumlimits, intlimits, namelimits, fleqn]{amsmath}
\usepackage{natbib}
\usepackage{bm}
\usepackage{hyperref}
\usepackage{siunitx}
\usepackage{enumitem}

\ifdefined\drafting
	\usepackage[left=0.25in,right=1.8in,head=21.0pt]{geometry}
\else
\fi
\usepackage[dvipsnames,svgnames,x11names]{xcolor}
\ifdefined\showcomments{}
	\usepackage[authormarkup=none,commentmarkup=todo,todonotes={textsize=scriptsize,textwidth=1.6in}]{changes}
	\makeatletter \def\@captype{figure} \makeatother 
\else
	\usepackage[final]{changes}
\fi
\newcommand{\markhigh}[1]{\bgroup\markoverwith				
  {\textcolor{#1}{\rule[-.5ex]{2pt}{2.5ex}}}\ULon}

\definechangesauthor[name=Frank Hohls, color=red]{FH}
\definechangesauthor[name=Slava Kashcheyevs, color=red]{VK}

\newcommand{\mulever}[1]{\alpha^{(\mu)}_#1}

\graphicspath{ {./figures/} }

\begin{document}

\title{Controlling the error mechanism in a tunable-barrier non-adiabatic charge pump by dynamic gate compensation}

\date{\today}

\author{Frank \surname{Hohls}}\email[Mail to:
]{frank.hohls@ptb.de}\affiliation{Physikalisch-Technische
Bundesanstalt, Bundesallee 100, 38116 Braunschweig, Germany.}

\author{Vyacheslavs Kashcheyevs}\affiliation{Departments of Physics,
University of Latvia, Riga LV-1002, Latvia.}

\author{Friederike \surname{Stein}}\affiliation{Physikalisch-Technische
Bundesanstalt, Bundesallee 100, 38116 Braunschweig, Germany.}

\author{Tobias \surname{Wenz}}\affiliation{Physikalisch-Technische
Bundesanstalt, Bundesallee 100, 38116 Braunschweig, Germany.}

\author{Bernd \surname{Kaestner}}\affiliation{Physikalisch-Technische
Bundesanstalt, Abbestr. 2-12, 10587 Berlin, Germany.}

\author{Hans W. \surname{Schumacher}}\affiliation{Physikalisch-Technische
Bundesanstalt, Bundesallee 100, 38116 Braunschweig, Germany.}

\begin{abstract}

Single-electron pumps based on tunable-barrier quantum dots are the most promising candidates for a direct realization of the unit ampere in the recently revised SI: they are simple to operate and show high precision at high operation frequencies. The current understanding of the residual transfer errors at low temperature is based on the evaluation of backtunneling effects in the decay cascade model. This model predicts a strong dependence on the ratio of the time dependent changes in the quantum dot energy and the tunneling barrier transparency. Here we employ a two-gate operation scheme to verify this prediction and to demonstrate control of the backtunneling error. We derive and experimentally verify a quantitative prediction for the error suppression, thereby confirming the basic assumptions of the backtunneling (decay cascade) model. Furthermore, we demonstrate a controlled transition from the backtunneling dominated regime into the thermal (sudden decoupling) error regime. The suppression of transfer errors by several orders of magnitude at zero magnetic field was additionally verified by a sub-ppm precision measurement.

\end{abstract}

\maketitle

\section{Introduction}
\label{sec:Intro}

Single-electron pumps based on tunable barrier quantum dots (QDs)~\cite{kouwenhoven1PBI, Ono2003a} show much promise as quantum current standard~\cite{Kaestner2015,Stein2017,Giblin2019,Giblin2020} within the revised SI-system that came into effect on May 20th, 2019. Since then, the value of the elementary charge $e$ in SI units is defined as $1.602176634\times10^{-19}$\,As without uncertainty. \added[id=FH]{As a consequence, any realization of a single-electron current source producing an exact integer multiple of the current $ef$ when operated with frequency $f$ is a direct realization of the SI unit of electrical current, the ampere. Among the many possible realizations scrutinized, e.g. normal metal, hybrid and superconducting turnstiles or pumps, surface acoustic wave based transport, quantum phase slip devices, reviewed extensively in Ref.~\cite{Pekola2013}, the tunable barrier QD based single-electron pumps are closest to application in current metrology~\cite{Giblin2019}.
In these devices, }modulation of the barriers defining the QD allows \added[id=FH]{for} the clocked transfer of electrons that results in an overall dc-current of the form $I=\left\langle n\right\rangle e f$, where $\left\langle n\right\rangle$ is the average number of transferred electrons per cycle, $e$ is the electron charge and $f$ the barrier modulation frequency. 
 Electron pumps  have been shown to operate at high frequencies~\cite{fujiwara1, blumenthal2007a, Kaestner2007c, fujiwara2008, jehl2013}, generating macroscopic dc currents with strong noise suppression~\cite{maire2008}.
They have been increasingly employed as on-demand electron sources in various circuits~\cite{feve2007, kaestner2010c, Leicht2011, fricke2011, yamahata2011, Hohls2012, Bocquillon2013, Fletcher2013, Fricke2013, Mirovsky2013, Fricke2014,Ubbelohde2014,Fletcher2019,Freise2020}. 

Non-adiabatic single gate modulation of QDs is a robust mode of operation and was investigated extensively~\cite{Kaestner2007c, giblin2010a, Hohls2012, giblin2012, Fletcher2012, yamahata2014,Stein2015,Stein2017,Giblin2019}.
For GaAs-based single-electron pumps a magnetic field is applied to enhance the pumping accuracy~\cite{Wright2008,kaestner2009a,giblin2012} with demonstrated accuracies down to $0.16\,\,$ppm, limited by uncertainties in the measurement setup~\cite{Stein2015,Stein2017,Giblin2019,Giblin2020}.
Generally, underlying error mechanisms are explained by backtunneling errors using the decay cascade model~\cite{Kashcheyevs2010}.
In this model, the plunger-to-barrier ratio $\Delta_\text{ptb}$ {quantifies the shift of the quantum dot energy level relative to the tunnel barrier closing speed.}
It has been predicted that a reduction of $\Delta_\text{ptb}$ leads to an improvement in pump accuracy, if the temperature is sufficiently low~\cite{Fricke2013,yamahata2014,Kashcheyevs2014,Yamahata2021}. 
In this paper, we demonstrate the predicted suppression of the dominating backtunneling error using a dynamic compensation scheme, in which the second gate effectively slows down the shift of the energy level. We derive a quantitative prediction for this effect from geometric arguments and verify this prediction experimentally for a GaAs-based single-electron pump without applied magnetic field.
Furthermore, we investigate the crossover into a regime, where thermal errors dominate the pump performance, as the shift of the QD energy level is suppressed during the closure of the barrier. For optimized parameters we observe a suppression of backtunneling errors by several orders of magnitude, experimentally verified with sub-ppm uncertainty by a precision measurement of the \added[id=FH]{current generated by the single-electron pump}.

\section{Experimental setup}
\label{sec:exp}

\begin{figure}
	\includegraphics[scale=1]{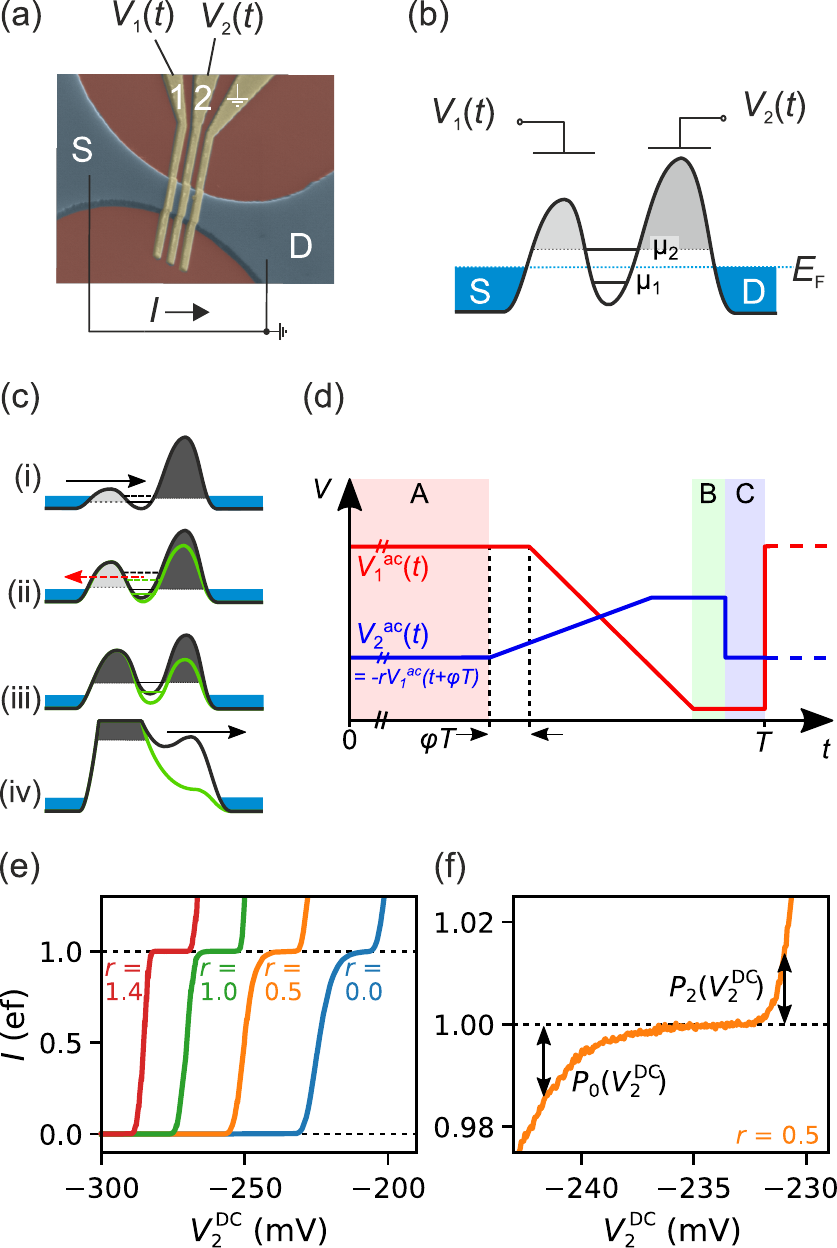}
	\caption{(a)~Scanning electron micrograph of a device similar to the one used for measurements.
	A quantum dot is formed between gates~1 and~2 and connected to source (S) and drain (D).
	The third gate is grounded.
	(b)~Schematic of the potential along the channel.
	Gate voltages $V_1(t)$ and $V_2(t)$ are modulated.
	(c)~\added[id=FH]{Black lines:} Phases of the basic single electron pumping process with constant $V_2$, i.e. $r=0$:
	(i)~Loading, (ii)~backtunneling, (iii)~decoupling, and (iv)~unloading. \added[id=FH]{Green lines: schematic potential profiles for $r>0$.}
	(d)~Waveform applied to both gates, where $V_2^\mathrm{RF}(t)=-r V_1^\mathrm{RF}(t+\varphi T)$.
	(e)~First plateau of quantized current for various gate amplitude ratios $r$.
	Quantization improves with increasing $r$.
	(f)~Zoom in on plateau.
	$P_0$ indicates probability of failure to capture the first electron.
	$P_2$ indicates probability to capture the second electron.}
	\label{fig:device}
\end{figure}

The device used in this work is shown in Fig.~\ref{fig:device}(a).
It is similar to tunable barrier QD pumps used in previous studies~\cite{Leicht2011,Stein2015,Stein2017}.
From a GaAs/AlGaAs heterostructure with carrier density of $2.1 \cdot 10^{15}/\, \text{m}^2$ and mobility of $280\, \text{m}^2/$Vs a channel of approximately $680$\,nm width was etched.
Two Ti/Au finger gates were used to form a quantum dot, a third gate was grounded.
The distance between gate centers was $250\,$nm.
The gates were driven by custom waveforms, generated by an arbitrary waveform generator (AWG) with a sampling rate of $12$\,GS/s and filtered using $1870$\,MHz low pass filters.
DC and AC signals were combined using bias-tees at room temperature~\footnote{High frequency gate control setup is described in detail in the supplemental of~\cite{Wenz2019}}.
A schematic potential landscape of a static quantum dot is shown in Fig.~\ref{fig:device}(b).

The device was placed in a dilution refrigerator with base temperature $T_{\mathrm{cryo}}=100 \,\mathrm{mK}$. 
All measurements were taken at zero magnetic field~\footnote{Perpendicular magnetic field is often applied to improve pump performance in GaAs pumps~\cite{kaestner2009a}}.
The generated electrical current was measured with two ultrastable low-noise current amplifiers (ULCA)~\cite{Drung2015} connected to source and drain, allowing for a relative noise floor of about $\delta I_{\mathrm{rms}}/I \approx 3 \cdot 10^{-4}$ in a measurement time of about $300\,$ms per point for a current of $I \approx e\cdot 85.4 \,\mathrm{MHz} = 13.5 \,\mathrm{pA}$. Two-dimensional pumping maps~\cite{Leicht2010} in the $V_1^\text{DC}$-$V_2^\text{DC}$-plane were taken and used to select a $V_1^\text{DC}$ working point, see appendix \ref{supp:pumpmap}.  

The standard operation principle of a non-adiabatic single-electron pump is shown in Fig.~\ref{fig:device}(c) \added[id=FH]{with the real-space potential sketched by a black line}.
Only $V_1(t)$ is modulated, while $V_2$ remains static.
(i)~The entry barrier $V_1$ is low and electrons are loaded into the quantum dot from the source.
(ii)~$V_1$ is lowered, raising the barrier and the quantum dot states.
This starts the decay cascade process~\cite{Kashcheyevs2010,Kaestner2015} in which superfluous electrons tunnel back to source.
(iii)~The quantum dot is decoupled and the number of electrons $n$ on the dot remains constant.
(iv)~Remaining electrons are unloaded to the drain side.
By repeating this process with frequency $f$, a current $I=\left\langle n\right\rangle ef$ is produced.
The average number $\left\langle n\right\rangle$ of electrons captured in the backtunneling phase can be controlled by changing the DC component of $V_2$.
A measurement of current for modulation of $V_2$ only is shown as blue curve in Fig.~\ref{fig:device}(e) ($r = 0$).
A plateau can be observed at $I=1ef$.

\section{Pumping scheme for suppression of backtunneling error}
\label{sec:scheme}

In the following we will show and illustrate how a modulation of $V_2(t)$ can be used to improve pump performance in a non-adiabatic single electron pump.
The dominating error mechanism is backtunneling~\cite{Kashcheyevs2010,Kaestner2015,Giblin2019}, which should therefore be reduced. 
To achieve this, we lower the rate of change of the QD energy level  during decoupling from the source (and thereby  reduce $\Delta_\mathrm{ptb}$~\cite{Fricke2013,yamahata2014,Kashcheyevs2014,Yamahata2021}) by applying a phase shifted inverse fraction of the waveform $V_1^\mathrm{RF}(t)$ to $V_2^\mathrm{RF}(t)=-r V_1^\mathrm{RF}(t+\varphi T)$.
The waveform $V_1^\mathrm{RF}(t)$ consists of three sections: a plateau for $7.9\,$ns, a ramp of $3.3\,$ns and another plateau of $3.3$\,ns, resulting in a repetition rate of $84.5$\,MHz. \added[id=FH]{Fig.~\ref{fig:device}(d) shows the time dependent signals applied to the two gates. The phase shift $\varphi=0.042$ and thus the time shift $\Delta t=\frac\varphi T\approx \SI{0.5}{ns}$ between $V_1(t)$ and $V_2(t)$ were chosen sufficiently large to allow settling of voltages in phase C, thus avoiding capture of electrons from drain.} The peak-to-peak amplitude of the signal $V_1^\mathrm{RF}$ is $165$\,mV. \added[id=FH]{The modified potential evolution caused by the additional voltage applied to the second gate for $r>0$ and the resulting reduced change of the QD energy is shown schematically by the green line in Fig.~\ref{fig:device}(c)}.
By increasing the ratio $r$ of the RF amplitudes applied to the gates, the steepness of the steps leading to and from the $1ef$ plateau can be increased, cf. Fig.~\ref{fig:device}(e).
This is associated with increased pumping accuracy on the plateau.

\begin{figure*}%
	\centering \includegraphics[scale=0.97]{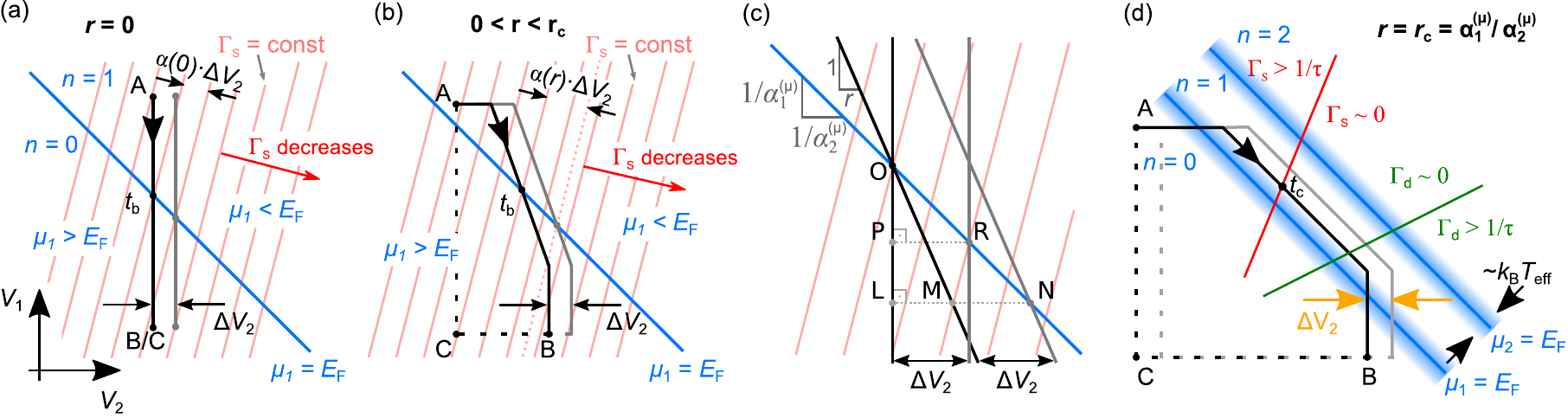}%
	\caption{Pumping process shown on the stability diagram.
		Blue line indicates resonance of first electron state with the leads.
		Red lines indicate levels of constant tunneling rate and are spaced equidistant on logarithmic scale $\ln \Gamma_s$\added[id=FH]{; the rate decreases in the direction indicated by the red arrow}.
		The black and gray lines indicate RF voltage paths (gray shifted by $\Delta V_2$).
		Points A,B,C are associated with intervals of constant voltage in the gate voltage waveforms, see Fig.~\ref{fig:device}(d), dashed lines are fast transitions.
		(a)~$r=0$: Backtunneling of electrons sets in at $t_b$.
		The backtunneling rate can be reduced by $\Delta(\ln \Gamma_s)=\alpha(0)\cdot\Delta V_2$ by shifting the path (gray) to the right by some voltage $\Delta V_2$.
		(b)~$r>0$: Backtunneling of shifted gray path is stronger reduced than in (a) since more lines of $\Gamma_s$ are covered with the same shift of $\Delta V_2$.
		(c)~Zoom into relevant region with triangles for geometric derivation of $\alpha(r)$.
		(d)~For $r = \mulever{2}/\mulever{1}$ the RF voltage path is parallel to the $\mu_i$ lines throughout the capture phase (i.e. closure of the barrier to source, indicated by crossing with the red line). The capture probability is determined by the paths distance to the $\mu_i = E_F$ resonance lines and by the effective temperature broadening of the source Fermi function, indicated by the washed broadened blue lines. Note: here the pumping process is adiabatic and the direction of loading/unloading is controlled by the barrier transparencies $\Gamma_{s/d}$ during crossing of the resonance line; the QD is coupled to source above the red line during loading and to drain below the green line during unloading. 
		}%
	\label{fig:compensationsketch}%
\end{figure*}

\subsection{Reduction of backtunneling error by compensation}

The reduction of backtunneling error can be understood within a simple geometric picture shown in Fig.~\ref{fig:compensationsketch} and explained below.
The average number of captured electrons in the vicinity of the first plateau is modeled as 
\begin{align} \label{eq:Igeneral}
  \left\langle n \right\rangle = I(V_2^{\text{DC}})/ef =  1 - P_0(V_2^{\text{DC}}) + P_2(V_2^{\text{DC}}) \, 
\end{align}
where $P_0$ and $P_2$ are the probabilities to transfer $0$ or $2$ electrons in a given cycle, respectively, see Fig.~\ref{fig:device}(f). $P_0$ decreases and $P_2$ grows as a function of $V_2^{\text{DC}}$; in the decay cascade regime $P_2(V_2^{\text{DC}})$ is much steeper and the main error on the plateau is the backtunneling of the last electron,  
$P_0 \gg P_2 \approx 0$~\cite{Yamahata2021}.
The sharpening of the steps with $r>0$ is thus associated 
with a sharper dependence of $P_0$ on $V_2^{\text{DC}}$.

A qualitative picture and a quantitative prediction for the accuracy improvement can be understood considering the RF voltage path in the stability diagram of the pump as shown \added[id=FH]{in} Fig.~\ref{fig:compensationsketch}. The blue line $\mu_1(V_1,V_2)-E_F=0$ indicates the zero chemical potential for addition of the first electron relative to the Fermi level $E_F$ of the source lead for given instantaneous values of the gate voltages $V_{1,2}$. 
For $r=0$, as shown in Fig.~\ref{fig:compensationsketch}(a), the voltage path follows a black straight line parallel to the $V_1$-axis from the loading phase at A [Fig.~\ref{fig:device}c(i)] to the ejection phase at B/C [Fig.~\ref{fig:device}c(iv)].
Once $\mu_1>E_F$, backtunneling can occur [Fig.~\ref{fig:device}c(ii)].
The probability $P_0$ to loose the electron is~\cite{Fricke2013,Kashcheyevs2014,Kaestner2015},
\begin{equation}
  P_0=1 - e^{ - \int_{t_b} \! \Gamma_s(t) \mathrm{d}t } \approx 1 - e^{ -\Gamma_s(t_b) \tau } \approx \Gamma_s(t_b) \tau \, ,  \label{eq:P0expr}	
\end{equation}
where $\Gamma_s$ is the backtunneling rate, $t_b$ is the time at which $\mu_1=E_F$ and $\tau$ is the timescale for the reduction of $\Gamma_s$ due to the time-dependence of $V_1(t)$.
We assume an exponential dependence of the backtunneling rate on gate voltages $V_1$ and $V_2$, in particular 
\begin{equation}
    \Gamma_s(t_b) \propto e^{-\alpha V_2^\textrm{DC}}, \label{eq:Gamma_s}
\end{equation} with the parameter $\alpha$ determining the steepness of the quantisation steps as function of $V_2^\textrm{DC}$ via Eqs.~\eqref{eq:Igeneral} and \eqref{eq:P0expr}~\cite{fujiwara2008,Kaestner2015}.
A shift of $V_2^\textrm{DC}$ by $\Delta V_2$ moves the pumping path (new position marked in gray), 
and the corresponding change in $\Gamma_s$ (and hence in $P_0$) can be estimated geometrically by measuring the shift of the resonance crossing point $t_b$ with respect to the 
level lines of $\Gamma_s(V_1,V_2)$, equidistant on a logarithmic scale (shown in red).

At $r>0$, the pumping path in the $(V_2,V_1)$ stability diagram is tilted \added[id=FH]{due to the dynamic change of $V_2(t)$ introduced by $r>0$} as shown in Fig.~\ref{fig:compensationsketch}(b). The same shift $\Delta V_2$ causes a stronger reduction in backtunneling
for $r>0$ compared to \added[id=FH]{$r = 0$} as follows from comparing the number of level lines along the (blue) resonance lines. This leads to the sharpening of the current steps observed experimentally in Fig.~\ref{fig:device}(e). Combining the relevant parts of Figs.~\ref{fig:compensationsketch}(a) and (b) in Fig.~ \ref{fig:compensationsketch}(c) allows for a simple geometric derivation,
\begin{align}
  \frac{\alpha(r\!=\!0)}{\alpha(r)} = \frac{OR}{ON}=1-\frac{\mulever{2}}{\mulever{1}} r \, ,
  \label{eq:1overr}
\end{align}
where the QD lever arm factors, $\mulever{1}$ and $\mulever{2}$ for $V_1$ and $V_2$ respectively, determine the slope of the constant energy line ($OR$)~\footnote{Two similar triangles $OPR$ and $OLN$ can be formed by drawing $PR$ and $LN$ parallel to $V_2$ axis, therefore $OR/ON=PR/LN$. Since $PR=MN=\Delta V_2$, hence $PR/LN=MN/LN=1-(LM/LN)$. From the definition of slopes for $OM$ and $ON$ we have $LM=r \, LO$ and $\mulever{1} LO = \mulever{2} LN$, which gives Eq.~(4)}.
Equation \eqref{eq:1overr} predicts a linear reduction of the width $1/\alpha(r)$ of the quantization step with the amplitude ratio $r$.
This dependence extrapolates to zero at $r=r_c =\mulever{1}/\mulever{2}$
determined by the ratio of the QD \added[id=VK]{energy} lever arm factors.

\added[id=VK]{The two-gate linear compensation of energy shifts for tuning the non-adiabatic single-parameter pumping and the robust quantitative prediction \eqref{eq:1overr} constitute the main proposition of our work. We stress that despite the apparent simplicity of the geometric derivation,  the linear relation is non-trivial and directly testable, sidestepping the need to determine lever arms  for barrier sensitivity to gate voltages (see Appendix~\ref{supp:lever-arm}).}

In order to probe the error reduction relation  \eqref{eq:1overr} we model the \added[id=VK]{pumping current} plateau by Eq.~\eqref{eq:Igeneral}
with 
\begin{align}
      P_0  & =  e^{- \alpha [ V_2^{\text{DC}} -
        V^{(1)} ]} \label{eq:P0fit} \, ,\\
        P_2  &=  e^{- \alpha' [ 
        V^{(2)} - V_2^{\text{DC}} ]}
        \label{eq:P2fit}
\end{align}
treating $\alpha$, $\alpha'$,  $V^{(1)}$ and $V^{(2)}$ as fitting parameters~\cite{Kashcheyevs2014}.

The fitting form Eq.~\eqref{eq:P0fit} for the probability 
to miss an electron follows from the backtunneling (decay cascade) mechanism described above, 
Eqs.~\eqref{eq:P0expr} and \eqref{eq:Gamma_s}.
The same mechanism for the capture of the (unwanted) second electron
predicts~\cite{Kashcheyevs2010} a steep double-exponential $V_2^{\text{DC}}$-dependence of $P_2$, however, at low enough $P_2$
 other error mechanisms~\cite{Kashcheyevs2012a,Fricke2013,yamahata2014} take over, with rare thermal errors following
 the exponential ansatz~\eqref{eq:P2fit}. 
\added[id=VK]{In particular, a recent theoretical study exploring the crossover between thermal and backtunneling errors using an isothermal multi-level QD model~\cite{Yamahata2021} is consistent with Eqs.~\eqref{eq:P0fit} and \eqref{eq:P2fit}, see Appendix~\ref{supp:lever-arm} for details.}

\subsection{Fully compensated regime}
\label{sec:compensated}

It is also instructive to conceptualize the 
device operation in the fully compensated ($r=r_c$) regime. The corresponding pumping contour is sketched in Fig.~\ref{fig:compensationsketch}(d).
The straight segment with the slope $r_c$ runs parallel to the constant energy line, making a single-parameter quantized operation~\cite{Blumenthal2007,Kaestner2007c} impossible. 
Our choice of $\varphi=0.042 \not = 0$ makes the device at $r=r_c$ a two-parameter tunable-barrier pump~\cite{Ono2003a} operating in the sudden decoupling limit~\cite{Kashcheyevs2012a,Fricke2013,yamahata2014,Yamahata2021}\added[id=VK]{: the chemical potentials $\mu_{1,2}$ remain constant with respect to $E_F$ in the source as the tunneling rate $\Gamma_S$  is quickly tuned from sufficiently large (adiabatic) to negligible (decoupled), thus producing a thermal (generalized grand canonical) distribution for the number $n$ of  electrons captured at around the time moment marked $t_c$ in Fig.~\ref{fig:compensationsketch}(d). 
The local equilibrium distribution of $n$ does not depend on the position $t_c$, as long as the latter remains on a line parallel to $\mu_{1,2}=\text{const}$.
From a standpoint of classifying quantized current pumps~\cite{Kaestner2015}, this fully compensated regime can be described as an adiabatic pumping scheme~\cite{Ono2003a,jehl2013} as it can yield $\langle n \rangle \to 1$ at arbitrary slow operation. Yet at the relevant high repetition frequencies (which make the decoupling transition possible) the dominating errors are controlled only by a short segment  near $t_c$  and not by the global geometry of the pumping contour as typically associated with the adiabatic pumping~\cite{Pothier1992} so that modelling and approximation techniques for a more generic single-parameter \cite{Kaestner2007c} (which can be two-gate, as in the pioneering Ref.~\onlinecite{blumenthal2007a}) non-adiabatic pumps \cite{Kaestner2015} can be employed.}

In the fully compensated regime $(r = r_c)$ the effects of backtunneling are  eliminated, and the  grand canonical distribution frozen by the sudden decoupling makes one expect~\cite{Fricke2013,yamahata2014,Yamahata2021}
\begin{equation}
    \alpha=\alpha' =\frac{\mulever{2}}{k_{B} T_{\text{eff}}}
    \label{eq:thermalalpha}
\end{equation} 
with some effective temperature $T_{\text{eff}}$ for both extra and missing electron errors.
\added[id=VK]{This can be understood geometrically from the sketch in  Fig.~\ref{fig:compensationsketch}(d) as $\Delta V_2$ simply shifts the linear part of the pumping contour (marked in sold black) relative to the resonance lines (marked in solid blue) regardless of the position of $t_c$. Equation \eqref{eq:thermalalpha} for the fully compensated limit is also consistent with a uniform temperature model~\cite{Yamahata2021} as shown in Appendix~\ref{supp:lever-arm}.}

\section{Measurement results}
\label{sec:results}

\begin{figure}
	\includegraphics[scale=1]{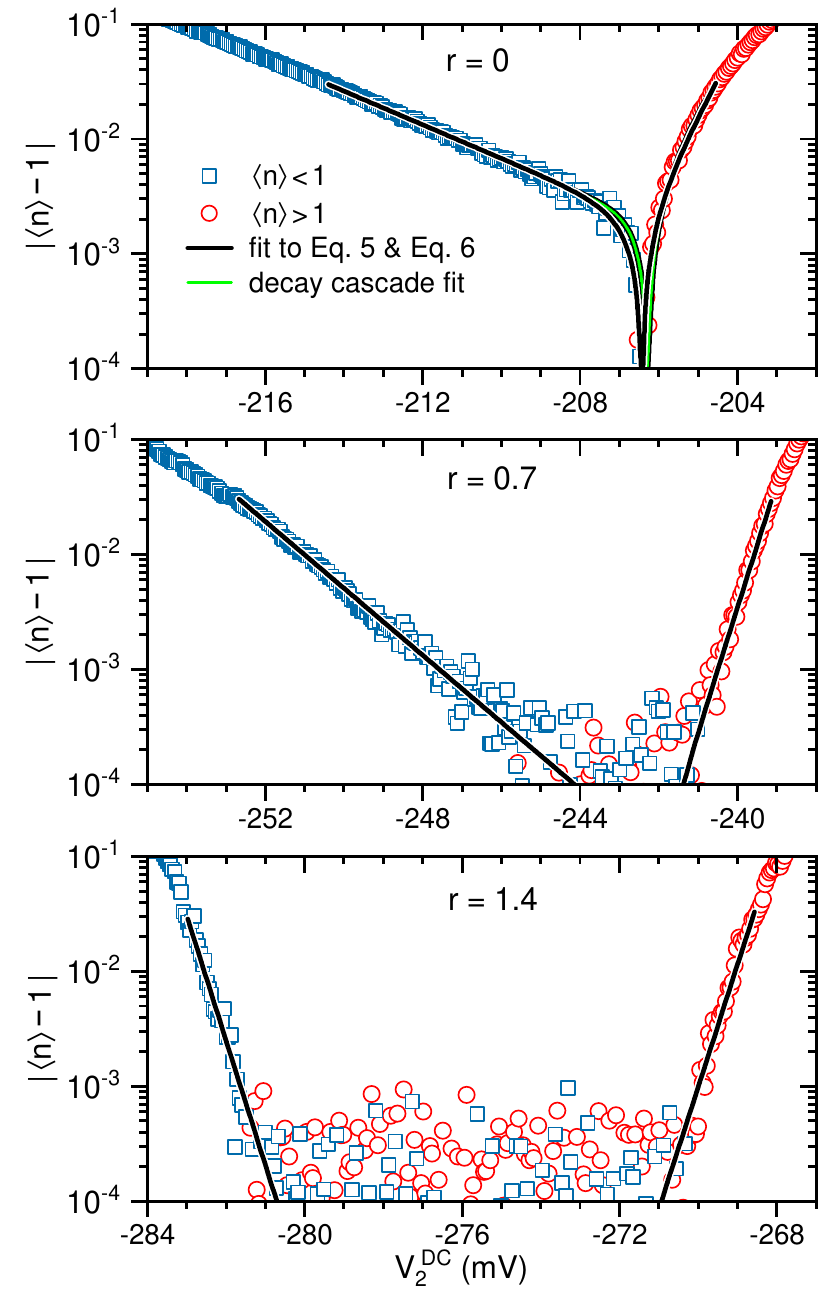}
	\caption{Capture probability $\left\langle n \right\rangle$ displayed as deviation from $1$ for $r = 0$, $0.7$ and $1.4$.
	Data fitted to $\left\langle n \right\rangle - 1 = -P_0 + P_2$ with $P_{0/2}$ from Eq.~\ref{eq:P0fit} and Eq.~\ref{eq:P2fit} (black lines). Blue squares and red circles are used to distinguish measured data with $\left\langle n \right\rangle < 1$ from $\left\langle n \right\rangle > 1$. For $r=0$ a fit of the original decay cascade function~\cite{Kashcheyevs2010} is shown for comparison (green).}
	\label{fig:Fits}
\end{figure}

\subsection{Suppression of backtunneling errors}

We will now analyze the experimental data with respect to the expected exponential behavior of the error probability $P_0(V_2^{\text{DC}})$ and additionally also examine $P_2(V_2^{\text{DC}})$. Figure~\ref{fig:Fits} shows $\left|\left\langle n \right\rangle -1\right| = \left|- P_0 + P_2\right|$ for different values of the compensation ratio $r$ with logarithmic mapping.  In this presentation of data the left part is dominated by $P_0$ and nicely reveals the expected exponential dependence $P_0\propto e^{-\alpha V_2^{\text{DC}}}$ with increasing slope $\alpha$ for increasing $r$. The right part is dominated by $P_2$ and shows for low $r$ a much steeper slope than $P_0$ as expected for a single-electron pump in the backtunneling error dominated regime~\cite{Kashcheyevs2012a,Yamahata2021}. In between there is a regime dominated by the competition of $P_0$ and $P_2$ which for low $r$ is also resolved in the measured data.

The black lines show best fits of the model functions $P_0$ and $P_2$ introduced in Eqs.~\ref{eq:P0fit} and \ref{eq:P2fit} with free parameters $\alpha$, $\alpha'$, $V^{(1)}$, $V^{(2)}$ (see section \ref{supp:data-fitting} for method). For comparison a fit of the classic decay cascade formula, $\langle n \rangle =
\exp[ -e^{-\alpha V_2^{\text{DC}}+\delta_1}] +
\exp[ -e^{-\alpha V_2^{\text{DC}}+\delta_1+\delta_2}]$,
to $r=0$ data with three fitting parameters $\alpha$, $\delta_1$, $\delta_2$ is shown by the green line. We find that already for $r=0$ the model introduced in Eqs.~\ref{eq:P0fit} and \ref{eq:P2fit} fits the data as good as the classic decay form and thus will be used to evaluate the $r$-dependence of $\alpha$ for all compensation ratios $r$.

\begin{figure}
	\includegraphics[scale=1]{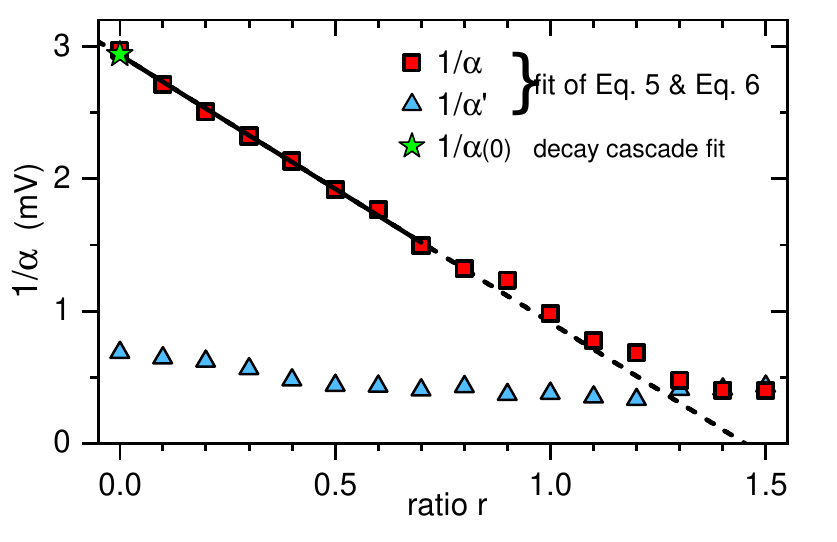}
	\caption{Fitting parameters $1/\alpha$ (squares) and $1/\alpha'$ (triangles) determined from fitting $P_0$ and $P_2$ to Eq.~\ref{eq:P0fit} and Eq.~\ref{eq:P2fit} as a function of gate amplitude ratio $r$. \added[id=FH]{The uncertainties of the parameter fit are smaller than the symbols size.} The green star shows $1/\alpha(0)$ determined from fitting the original decay cascade function at $r=0$. The black line shows a linear fit of $1/\alpha$ for $r\leq 0.7$ according to Eq.~\ref{eq:1overr}, the dashed line its extrapolation to \added[id=FH]{the horizontal intercept ($1/\alpha = 0$), where $r=r_c$}.}
	\label{fig:mainResult}
\end{figure}

The main result of this measurement, the $r$-dependence of the backtunneling error slope $\alpha$, is shown in Fig.~\ref{fig:mainResult} in reciprocal form. We immediately notice the linear suppression of $1/\alpha$ with increasing $r$ predicted in Eq.~\ref{eq:1overr}. This nicely confirms our geometric derivation of backtunneling error reduction by compensation~\footnote{The minimal pumping error $P_0+P_2$ additionally depends on the distance $\Delta V_2^\text{step}$ between the current steps, which is reduced slightly by increasing $r$, c.f. appendix \ref{supp:plateausize}; however, this effect influences the pumping accuracy much less than the strong reduction of $1/\alpha$}. Furthermore, we notice the saturation of $1/\alpha$ at large $r$, where other error mechanisms become visible due to the strong suppression of backtunneling error. The full compensation ratio $r_c$ is extracted by a linear fit of $1/\alpha(r)$, restricted to $1/\alpha(r) > 1/(2\alpha(0))$ to avoid including the saturation effect. The extracted $r_c = 1.45 > 1$ means that the effect of gate $1$ onto the QD energy is stronger than that of gate $2$, i.e. the QD is closer to gate $1$. This matches the \added[id=FH]{generally accepted }expectation for the asymmetric gating in the backtunneling phase as sketched in Fig.~\ref{fig:device}c(ii)\added[id=FH]{, see also Ref.~\cite{Yamahata2019}}.

We will now discuss the results for $r \rightarrow r_c$. Here we expect the backtunneling error to be suppressed, $1/\alpha \rightarrow 0$. This suppression reveals further electron loss error mechanisms in $P_0$ which have been masked by the backtunneling errors at lower $r$, now visible in measured $1/\alpha$-values lying above the extrapolation of the linear fit. Here we observe the expected transition to the thermal (sudden decoupling) regime~\cite{Fricke2013}. 
In this regime it is instructive to compare $\alpha$ to $\alpha'$, the slope parameter for an exponential onset of $P_2$ errors. We observe $\alpha \rightarrow \alpha'$ at $r \rightarrow r_c$ as now both the $P_0$ and the $P_2$ probability are governed by a grand canonical distribution with some effective temperature $T_{\text{eff}}$ (Eq.~\ref{eq:thermalalpha}). Note that the quantized transport mechanism now is a two-gate single QD pumping with the pumping contour encircling the first QD resonance that does not require non-adiabaticity to work~\cite{Kaestner2015,Ono2003a,jehl2013}.

Next we will examine the exponential slope parameter $\alpha'$ of fitting $P_2$ (Eq.~\ref{eq:P2fit}). Fig.~\ref{fig:mainResult} reveals a nearly constant $\alpha'(r) \approx \alpha'(r_c)$ for $r \gtrsim 0.5$. This experimentally justifies our reasoning in section \ref{sec:scheme} to describe $P_2$ by a simple exponential function, describing the capture of a thermally activated second electron due to an effective system temperature $T_\text{eff}$. 
The change of $\alpha'$ towards $r=0$ presumably mirrors the onset of additional second electron capture from the Fermi sea due to the decay cascade mechanism at large $1/\alpha$.

\begin{figure}%
	\includegraphics[scale=1]{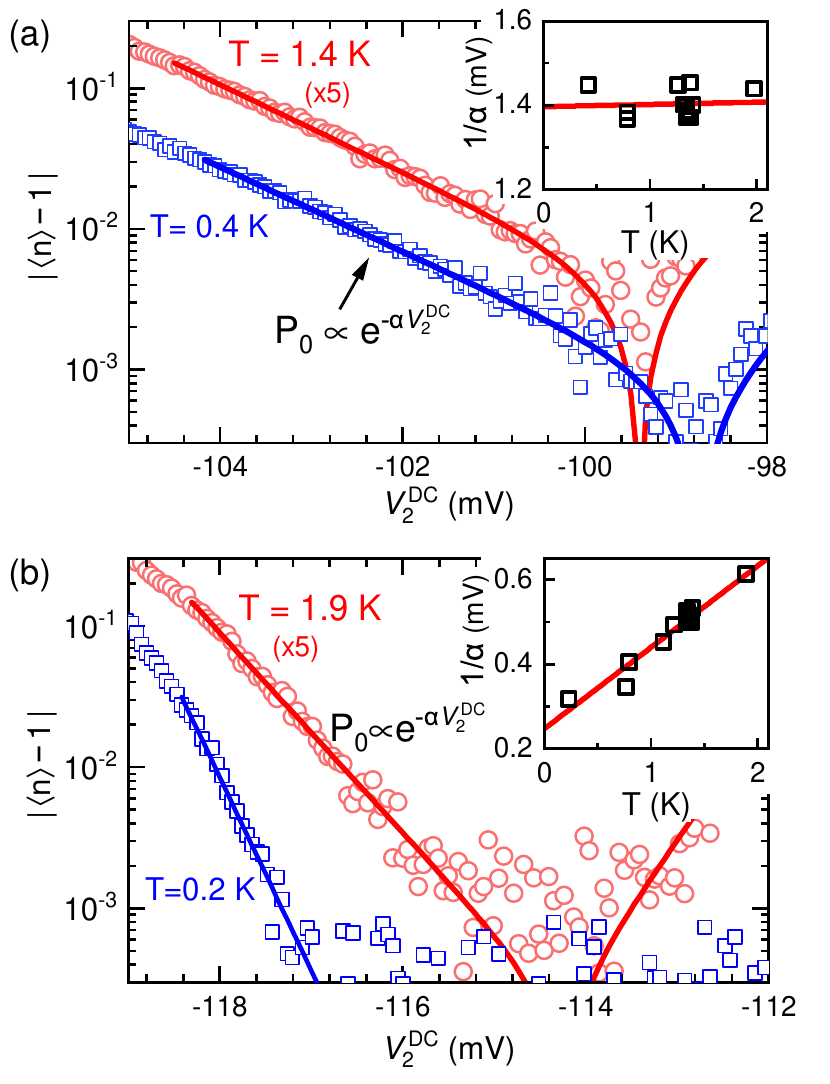}%
	\caption{Exemplary temperature dependence of the capture probability for low (blue squares) and high (red circles, data multiplied by 5) temperatures as indicated in the graphs for gate amplitude ratio (a) $r=0$ and (b) $r=1.2$ respectively. $T \equiv T_\text{cryo}$. Lines show fits to $\left\langle n \right\rangle - 1 = -P_0 + P_2$ with $P_{0/2}$ from Eq.~\ref{eq:P0fit} and Eq.~\ref{eq:P2fit}. The insets show the determined values of $1/\alpha$ as function of temperature (black squares), the line is a linear fit of $1/\alpha(T)$.
	Data were taken in a different sample cooldown with shifted gate voltage characteristic and slightly changed full compensation ratio $r_c\approx 1.2$.}%
	\label{fig:temperature}%
\end{figure}

\subsection{Temperature dependence}

The transition from a backtunneling dominated (decay cascade) error regime to a thermal (sudden decoupling) regime should also be visible in the dependence on the temperature $T_\text{cryo}$ of the cryostat hosting the device: for the backtunneling regime we expect no change of the exponential error slope $\alpha$ as long as backtunneling errors are much more common than thermal errors, i.e. as long as $1/\alpha \gg k_BT_\text{cryo}/\alpha_2^{(\mu)}$ (cf. Eq.~\ref{eq:thermalalpha}). In contrast, in the sudden decoupling regime \added[id=FH]{we expect} $1/\alpha = k_BT_\text{eff}/\alpha_2^{(\mu)}$ with $T_\text{eff} = T_\text{cryo}$ for sufficiently high temperatures and some saturation of $T_\text{eff}$ for $T_\text{cryo} \rightarrow 0$.

The result of such a temperature dependent measurement is shown in Fig.~\ref{fig:temperature} with (a) in the backtunneling regime ($r=0$) and (b) in the sudden decoupling regime ($r \approx r_c$) for temperatures $T_\text{cryo} \lesssim 2$~K. The inverse slope exponent $1/\alpha$ displayed in the insets is extracted by fitting $\left\langle n \right\rangle - 1 = -P_0 + P_2$ as before.
As expected we observe a negligible temperature dependence of $1/\alpha$ for $r=0$, whereas $1/\alpha(T)$ significantly changes with $T_\text{cryo}$ for $r\approx r_c$. 
This confirms our attribution of the different pumping error regimes to the athermal (backtunneling) and the thermal limit.

\subsection{Precision measurement}

\begin{figure}%
	\includegraphics[scale=1]{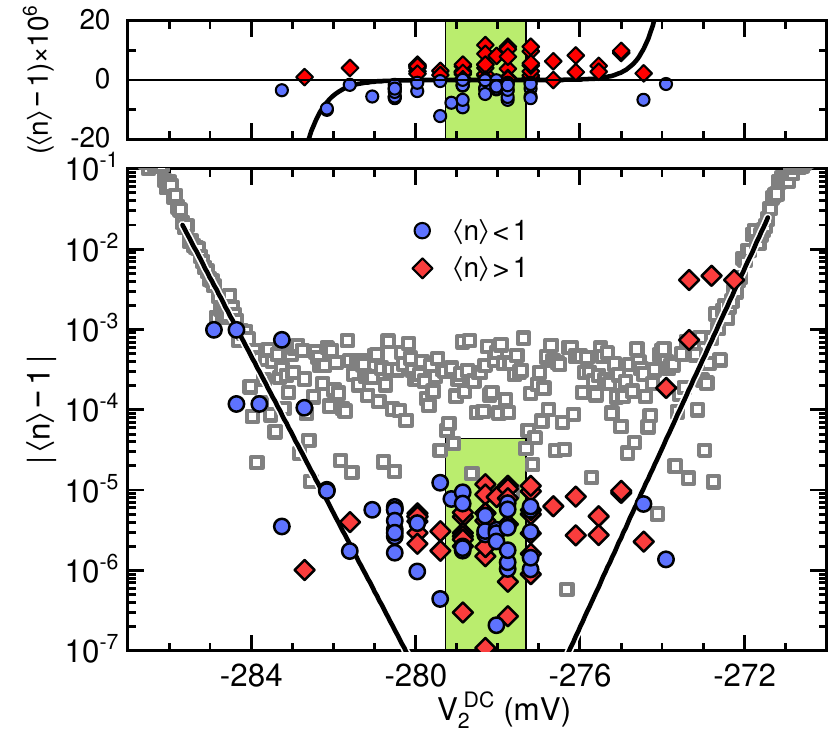}%
	\caption{Precision measurement of plateau center for $r=1.4$. Gray squares show result of an overview measurement with short (300 ms) integration time per point.  Lines show fit of $P_0$ and $P_2$ to this data. Filled symbols show 1 hour precision measurements with typical \SI{5}{\uA/A} uncertainty, see Ref.~\cite{Stein2017} for method, \added[id=FH]{color indicates sign of deviation, top graph shows precision data on linear scale}. Gate dependence has shifted compared to Fig.~\ref{fig:Fits} and shows further slight shifts on the time scale of the precision measurement, presumably due to trap activation by a discharging event. Averaging 45 values within the range indicated by the green rectangle results in $\left\langle n \right\rangle -1 = (0.65 \pm 0.75)\cdot 10^{-6}$.} 
	\label{fig:precision}%
\end{figure}

The suppression of backtunneling errors by two-gate operation immediately raises the question of application in electrical current metrology. The established procedure to achieve a metrologically relevant accuracy of less than 1 error in $10^6$ single electron transfers in devices based on GaAs/AlGaAs is the application of rather large magnetic fields of order 10~T\cite{giblin2012,Stein2015,Stein2017,Giblin2019}. The data for full compensation ($r=1.4$) in Fig.~\ref{fig:Fits} suggests achieving a relevant accuracy level without magnetic field. However, based on a 300 ms per point measurement, resolving $\left\langle n \right\rangle$ with a resolution of few parts in $10^4$, we cannot exclude other error mechanisms hidden in the noise. Thus we have applied a traceable precision measurement of the current using the setup and procedure described in Ref.~\cite{Stein2017} to examine the plateau region $\left\langle n \right\rangle \approx 1$. 
The result is shown in Fig.~\ref{fig:precision}. The precision measurements suggest an accuracy well below \SI{1}{\uA/A} at the center of the plateau, which can be compared to a minimal error probability of $P_0+P_2 \ge 4 \,\text{mA}/\text{A}$ for single gate operation ($r=0$). Averaging the precision measurements in the central region of the plateau shows an agreement with the desired $\left\langle n \right\rangle = 1$ within the measurement uncertainty of \SI{0.75}{\uA/A}. The larger uncertainty compared to Ref.~\cite{Stein2017} is caused by the 7-fold lower current level.

Before applying the demonstrated scheme for backtunneling error suppression by two-gate operation in practical current metrology, certain constraints must be overcome. Firstly, the repetition frequency in this experiment (84.4 MHz) and thus the generated current (13.5 pA) are comparably small, limiting the application range of this demonstration experiment. Application at higher frequencies requires either extensive engineering of the high frequency setup used to apply the driving signal or developing a simpler driving scheme which still offers sufficient error suppression. \added[id=FH]{We expect a significant reduction of backtunneling error also for simplified schemes with less precise voltage control at higher frequency, however, the achievable amount of error suppression at transfer rates near 1 GHz as used in demonstrations of sub-ppm accuracy in high magnetic field~\cite{Stein2015,Stein2017,Giblin2019,Giblin2020} still needs to be examined.}
\section{Conclusion}
\label{sec:Conclusion}

We have introduced and implemented a dynamic two-gate operation scheme to examine, to control and ultimately to suppress the dominating error process of non-adiabatic tunable barrier single-electron pumps. The observed dependence of the error suppression on our control parameter, the degree of compensation applied to the second gate, agrees well with the prediction for backtunneling errors derived from geometrical arguments. Thus our results underpin the assumptions underlying the widely used backtunneling (decay cascade) model. Furthermore, we could experimentally demonstrate the predicted~\cite{Kashcheyevs2012a,Fricke2013,Yamahata2021} transition from the backtunneling (decay cascade) to the thermal (sudden decoupling) error regime by dynamic reduction of the plunger action on the quantum dot energy and thereby of the plunger-to-barrier ratio $\Delta_{\text{ptb}}$. These results also open a route to improve the precision of GaAs-based single-electron pumps without need of large quantizing magnetic fields. Towards this aim the derived accuracy enhancement of several orders of magnitude was underpinned by a precision measurement with sub-ppm accuracy.

\section{Acknowledgements}

We thank Christoph Leicht and Thomas Weimann for help with the sample fabrication, Holger Marx and Klaus Pierz for providing the GaAs/AlGaAs heterostructure, Martin Götz for the calibration of the ULCAs and Thomas Gerster for critical reading of the manuscript. This work was supported in part by the Joint Research Projects e-SI-Amp (15SIB08) and SEQUOIA (17FUN04). These projects received funding from the European Metrology Programme for Innovation and Research (EMPIR) cofinanced by the Participating States and from the European Unions Horizon 2020 research and innovation programme. H.W.S acknowledges funding by the Deutsche Forschungsgemeinschaft (DFG, German Research Foundation) under Germany’s Excellence Strategy – EXC-2123 QuantumFrontiers – 390837967. V.K acknowledges funding by the Latvian Council of Science (project no. lzp-2018/1-0173).

\appendix

{
\section{Energy scales and lever arm factors\label{supp:lever-arm}}

Here we summarize an algebraic path to derive the main Eq.~\eqref{eq:1overr} and relate our work to the models available in the literature~\cite{Kashcheyevs2012a,Fricke2013,Kashcheyevs2014,Yamahata2021}.

Quantitative modeling of the current generated by tuneable barrier pumps typically requires the knowledge of  two types of lever arm factors for each gate involved, characterizing the shift in energy $\mu_1$ (`plunger function') and the change in escape rate $\Gamma_s$ (`barrier function') in response to each gate voltage, respectively.  The plunger and the barrier lever arms can be defined \cite{Kaestner2015} respectively as 
 $\alpha^{(\mu)}_{1,2}=- \partial{\mu_1}/\partial V_{1,2}>0$, and $\alpha^{(b)}_{1,2}=-\partial \log \Gamma_S/\partial V_{1,2}$.  The entrance barrier is closing  as $V_1$ is made more negative ($\alpha^{(b)}_1<0$) while the  gate $2$ has the opposite effect on the same barrier: the backtunnelling is enhanced as $V_2$ becomes more negative,  $\alpha^{(b)}_2>0$.

 For a finite compensation ratio $r>0$, the backtunneling stage [(ii) in Fig.~\ref{fig:device}(c)] is still controlled by a single time-dependent parameter  constrained by $r V_1(t)^{\text{RF}}+ \, V_2^{\text{RF}}(t) =\text{const}$, and the above definitions can be used to express the corresponding time dependence of $\mu_1(t)$ and $\Gamma_S(t)$, and hence the the triggering time for backtunneliling $t_b$ and the time constant for the closing of the barrier $\tau$ in Eq.~\eqref{eq:P0expr}.  This straightforward calculation~\cite{Kashcheyevs2012a} yields 
\begin{align}
\Delta_{\text{ptb}}(r)& =-d \mu_1/d (\log \Gamma_S) 
 =\frac{\alpha^{(\mu)}_1- r \alpha^{(\mu)}_{2}}{-\alpha^{(b)}_1+ r \alpha^{(b)}_{2}} \, ,  \label{eq:DeltaPTBr} \\
\alpha(r\!=\!0)& =\alpha^{(b)}_{2}+ \alpha^{(\mu)}_{2}/ \Delta_{\text{ptb}}(r\!=\!0) \, , \label{eq:alpha0}
\end{align}
as well as Eq.~\eqref{eq:1overr}.
We observe that  $\alpha(0)/\alpha(r)$, in contrast to $\Delta_{\text{ptb}}(r)$, is independent of the barrier sensitivity coefficients $\alpha^{(b)}_{1,2}$ and is linear in $r$.  We also observe that even though both  $1/\alpha(r)$ and 
$\Delta_{\text{ptb}}(r)$ vanish for $r \to r_c$, , the limit
$\lim\limits_{r\to r_c} \alpha(r) \, \Delta_{\text{ptb}}(r) = \alpha^{(\mu)}_{2}$ is finite.

A recent theoretical study by Yamahata \emph{et al.} \cite{Yamahata2021} has explored the crossover between different charge capture regimes in a multi-level quantum dot model assuming thermal equilibrium for all degrees of freedom except the confined charge. Their results on $I(V_2^{\text{DC}})$ have been expressed in terms of $g=\Delta_{\text{ptb}} /kT_0$  where the additional energy scale $kT_0$
 characterizes  the transition from tunneling (at $T < T_0$) to thermally-activated hopping (at $T >T_0$) as the electron escape mechanism. 
 Asymptotically, the model of Ref.~\onlinecite{Yamahata2021} is consistent with our Eqs.~\eqref{eq:Igeneral}, \eqref{eq:P0fit}, and \eqref{eq:P2fit}.  In particular, that model predicts a temperature-independent $\alpha(T)=\alpha(T\!=\!0)$ given by our Eq.~\eqref{eq:alpha0} for $kT < \Delta_{\star}$
 and 
 \begin{align} \label{eq:yamahtaalpha}
 \alpha(T\!=\!0)/\alpha(T) & =   kT/\Delta_{\star}  \, \text{ for } \, kT> \Delta_{\star} \, ,
 \end{align}
where $\Delta_{\star}^{-1}= \Delta_{\text{ptb}}^{-1}+(k T_0)^{-1}$.

For the excess-electron error the model expectation is 
\begin{align} \label{eq:yamahtaalphaprime}
    \alpha'(T) & = \alpha(T\!=\!0)\, \Delta_{\text{ptb}}/kT
\end{align}
at any $T$ and a sufficiently small $P_2$. The latter regime (where the exponential ansatz \eqref{eq:P2fit} for $P_2$ applies) is termed `fast decoupling' in Ref.~\onlinecite{Yamahata2021} as a new intermediate regime between the  `sudden decoupling' and `decay cascade' regimes. The latter were introduced in a QD-model agnostic way in Ref.~\onlinecite{Fricke2013} as the `thermal' and `athermal' limiting forms of the captured electron number distribution, respectively.

QD models that neglect intradot excitation \cite{Kashcheyevs2012a,Kashcheyevs2014} correspond to  $kT_0 \to \infty$ limit (no thermally-induced  hopping as a mechanism for electron escape back to the source). As expected, the reported results on $P_1$ \cite{Kashcheyevs2012a} and $I/(ef)$ \cite{Kashcheyevs2014} modelled by Eq.~\eqref{eq:Igeneral}  are consistent with Eq.~\eqref{eq:yamahtaalpha}  for  $\Delta_{\star} = \Delta_{\text{ptb}}$ and with Eq.~\eqref{eq:yamahtaalphaprime}.

For our compensation scheme, each value of $r$ corresponds to a certain $\Delta_{\text{ptb}}(r)$ described by Eq.~\eqref{eq:DeltaPTBr},  and   $\Delta_{\star}(r)\to \Delta_{\text{ptb}}(r) \to 0$ as $r \to r_c$. We have verified by an explicit calculation that the model of Ref.~\onlinecite{Yamahata2021} implies the following temperature dependence of $\alpha$:
\begin{itemize}[nosep]
    \item[(i)] The linear reduction in $1/\alpha$ dominated by the backtunelling error, Eq.~\eqref{eq:1overr},  fails at some $r_{\star}<r_c$ such that $\Delta_{\star}(r_{\star})= k T$ and the thermally-driven exponent~\eqref{eq:yamahtaalpha} becomes applicable instead.
    \item[(ii)] At $r=r_c$ Eq.~\eqref{eq:yamahtaalpha} and Eq.~\eqref{eq:yamahtaalphaprime} give $\alpha'(T)=\alpha(T)$ because at this compensation ratio $\Delta_{\star}/\Delta_{\text{ptb}}=1$ and $1/\alpha(T\!=\!0)=0$ [follows from our Eqs.~\eqref{eq:DeltaPTBr} and \eqref{eq:alpha0}], hence we recover the prediction of symmetric plateaus in the fully compensated regime, Eq.~\eqref{eq:thermalalpha} with  $T_{\text{eff}}=T$.
\end{itemize} 
}
\section{Fitting of data}
\label{supp:data-fitting}
The model functions Eq.~\ref{eq:P0fit} and Eq.~\ref{eq:P2fit} for the pumping errors $P_0$ and $P_2$ are fitted to the measured $\left\langle n \right\rangle -1 = - P_0 + P_2$ by nonlinear least square optimization with weighting. The approximations of $P_0$ and $P_2$ used are valid for $P_{0,2} \ll 1$. To capture a sufficient range of $\ln\left(P_{0,2}\right)$ for a reliable fit of exponential functions we include data with $P_{0,2} \leq 0.03$, where the upper limit $0.03$ is chosen as hundredfold of the measurement noise floor  $\left\langle I/ef \right\rangle_\mathrm{rms} \approx 3\cdot 10^{-4}$ determined for blocked single-electron pumping at more negative gate voltages $V_2^\mathrm{DC}$. \added[id=FH]{We tested that the central fitting result $1/\alpha$ is not sensitive to this choice, see the supplemental \cite{Supplemental} for this test.} We devised a special weighting of the data in the fit based on two criteria: (1) We want to mimic the effective weighting given when determining the steepness $\alpha$ of an exponential function $y=e^{-\alpha x}$ by linear regression of $\ln y$; an equally weighted nonlinear square fit of $y=e^{-\alpha x}$ would give excessive weighting of the exponential slope for maximum data values. (2) This weighting has to be bounded at low values to avoid excessive influence of the measurement noise floor. The weights $w_i$ of data points $y_i = \left(I_i/ef\right)-1$ are calculated by $1/w_i = 3\cdot 10^{-4}\sqrt{\max\left(\left|y_i\right|,10^{-3}\right)\left\langle 1/\max\left(\left|y_i\right|,10^{-3}\right)\right\rangle_i}$. See the supplemental \cite{Supplemental} for graph collecting all fits performed. The measured data used for fitting and all data derived and shown in this manuscript is available at \cite{Data_Hohls2022}.

\section{Plateau width}
\label{supp:plateausize}

\begin{figure}[h]%
	\includegraphics[scale=1]{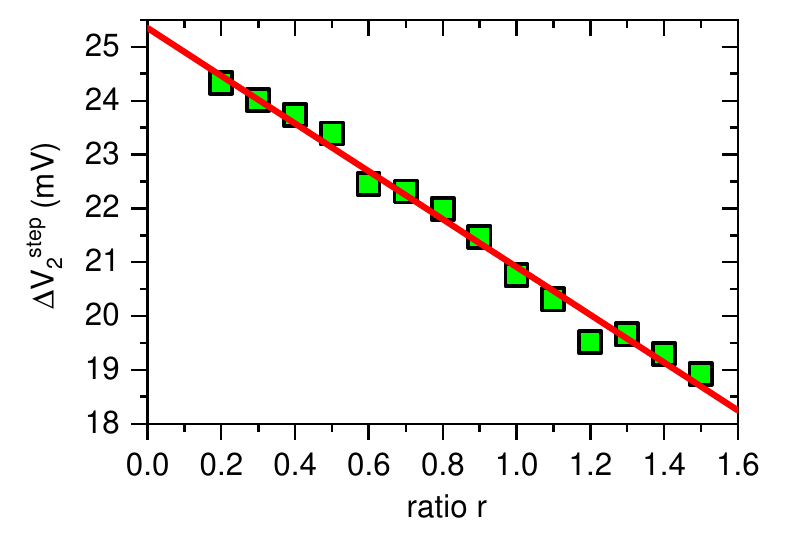}%
	\caption{Plateau width of $n=1$ plateau measured as voltage $V_2^\text{DC}$ distance from $\left\langle n \right\rangle=0.5$ to $\left\langle n \right\rangle=1.5$ (squares). The red line shows a linear fit to the data.}%
	\label{fig:plateausize}%
\end{figure}

The $n=1$ plateau width, measured as distance $\Delta V_2^\text{step}$ between the $n=1$ and $n=2$ step, is expected to follow
\begin{equation}
    \alpha(0) \, \Delta V_2^\text{step}(r) = \left (1 -
    \frac{r}{r_c} \right) \delta_2 +   \frac{\alpha(0) }{\mulever{2}} \left (\mu_2-\mu_1 \right ) 
    \frac{r}{r_c}\, ,
    \label{eq:plateausize}
\end{equation}
where $\delta_2 = (\mu_2-\mu_1  ) / \Delta_{\text{ptb}}(0)+\ln (\Gamma_2/\Gamma_1)$~\cite{Kaestner2015} is the dimensionless figure-of-merit of the decay cascade model~\cite{Kashcheyevs2010}.
The plateau width at $r=0$ is determined by the decay cascade mechanism, $\Delta V_2^\text{step}(0)= \delta_2/\alpha(0)$, and at full compensation by the addition energy, $\Delta V_2^\text{step}(r_c)=(\mu_2-\mu_1)/
\mulever{2}$, in accordance with the corresponding athermal and thermal limits of
the charge capture distribution~\cite{Fricke2013}, respectively. The experimentally determined plateau width $\Delta V_2^\text{step}$ as function of $r$ is shown in Fig.~\ref{fig:plateausize}, agreeing well to the expected linear $r$-dependence (Eq.~\ref{eq:plateausize}).

\section{Single-electron pumping map}
\label{supp:pumpmap}

\begin{figure}[h]%
	\includegraphics[width=8.5cm]{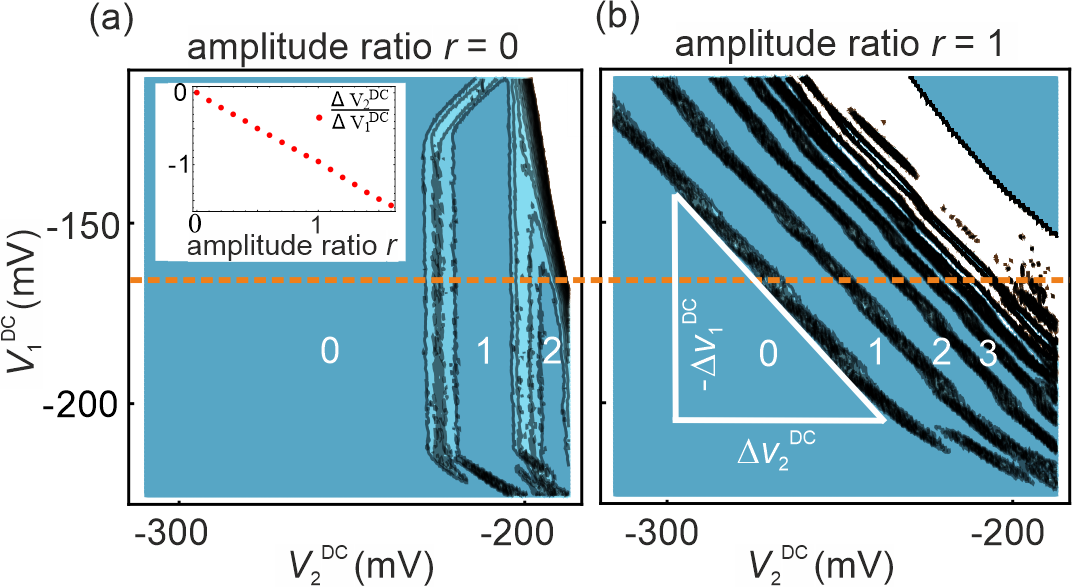}
	\caption{(a) Single-electron current map for single-parameter pamping ($r=0$); (b) map for two-gate pumping with RF source amplitude ratio $r=1$. Numbers $n$ mark the regions of $\left\langle n\right\rangle$ transferred electrons per cycle. 
	Inset shows linear dependence of tilt of the plateau regions vs. gate amplitude ratio $r$ with slope $-1$. Orange dashed line indicates position of cuts for further acquisition of the data analyzed in section \ref{sec:results}.}
	\label{fig:tiltResult}
\end{figure}

The symmetry of the RF signal lines used to apply the fast signals to gates 1 \& 2 can be tested by measuring the 2D pumping map, i.e. the pump current as function of $V_2^\text{DC}$ (x-axis) and $V_1^\text{DC}$ (y-axis). For single gate pumping ($r=0$, Fig.~\ref{fig:tiltResult}a) this map is well known and understood~\cite{Leicht2010} and shows a perpendicular constant current contour for the onset of the single electron pumping (invariance of $V_2^\text{DC}$ dependence on $V_1^\text{DC}$); a change in $V_1^\text{DC}$ leads to a shift in time $t_b$ for the onset of backtunneling~\cite{Kashcheyevs2012a}, but the instantaneous voltage $V_1(t_b)$ applied to the gate and thereby the dot potential are identical at this onset. Applying now RF signals to both gates with RF source amplitude ratio $r > 0$ changes the map, see Fig.~\ref{fig:tiltResult}b: the pumping onset current contour is tilted as a shift in time of the backtunneling onset also changes the instantaneous potential $V_2(t)$ at gate two. For a constant capture probability this has to be compensated by a shift of $V_2^\text{DC}$ such that both instantaneous voltages $V_{1}(t_b)$ and $V_{2}(t_b)$ at backtunneling onset stay constant. Thus we expect a shift $\Delta V_2^\text{DC} = -r\cdot \Delta V_1^\text{DC}$ for current onset when changing $V_1^\text{DC}$. Indeed this is observed as shown in the inset of Fig.~\ref{fig:tiltResult}a and confirms that the RF voltage ratio at the gates of the device is identical to the RF source amplitude ratio generated by the AWG. This also confirms the symmetry of the RF paths as any imbalance $\zeta = \eta_1/\eta_2 \neq 1$ of the RF transfer functions $\eta_{1,2}$ would have changed the tilt of the the pump map to $\zeta\cdot r$. The maps were also used to choose a $V_1^\text{DC}$ working point, centered on the $n=1$ plateau for $r=0$ and indicated by the orange dashed line in Fig.~\ref{fig:tiltResult}.

%

\end{document}